\documentclass[11pt]{iopart}
\usepackage{bm,amssymb,epsfig,cite,color}
\newcommand{\be}{\begin{equation}}
\newcommand{\ee}{\end{equation}}
\newcommand{\bea}{\begin{eqnarray}}
\newcommand{\eea}{\end{eqnarray}}

\newcommand{\p}{\partial}
\newcommand{\re}{\rm e}

\newcommand{\ri}{{\rm i}}
\def\affiliation{\address}

\def\nn{\nonumber\\}

\def\sgn{{\rm sgn}}
\def\eps{\epsilon}

\begin{document}
\title{On the spin-liquid phase of one dimensional spin-1 bosons}
\author{F.H.L. Essler$^1$, G.V. Shlyapnikov$^{2,3}$ and A.M. Tsvelik$^4$}
\affiliation{$^1$ The Rudolf Peiers Centre for Theoretical Physics,
  University of Oxford, 1 Keble Road, Oxford OX1 3NP, UK\\
$^2$ Laboratoire de Physique Th\'eorique et Mod\'eles
  Statistiques, Universit\'e. Paris Sud, CNRS, 91405~Orsay, France\\
$^3$Van der Waals-Zeeman Institute, Universitry of Amsterdam,
  Valckenterstraat 65/67, 1018 XE Amsterdam, The Netherlands\\
$^4$ Department of  Condensed Matter Physics and Materials
  Science, Brookhaven National Laboratory, Upton, NY 11973-5000, USA}
\date{\today}
\begin{abstract}
We consider a model of one dimensional spin-1 bosons with repulsive
density-density interactions and antiferromagnetic exchange. We
show that the low energy effective field theory is given by a
spin-charge separated theory of a Tomonaga-Luttinger Hamiltonian and
the O(3) nonlinear sigma model describing collective charge and spin
excitations respectively. At a particular ratio of the density-density
to spin-spin interaction the model is integrable, and we use the exact
solutions to provide an independent derivation of the low energy
effective theory. The system is in a superfluid phase made of singlet
pairs of bosons, and we calculate the long-distance
asymptotics of certain correlation functions.
\end{abstract}
\maketitle
\section{Introduction}
Spinor Bose gases exhibit a much richer variety of interesting
macroscopic quantum phenomena than spinless bosons. A mean field
analysis of three-dimensional spin-1 Bose-condensed gases
\cite{Ho,Ohmi} and exact diagonalization of the spin Hamiltonian 
for this system \cite{ueda}  
showed that the ground state can be either
ferromagnetic or spin-singlet, depending on the sign of the 
coupling constant for the exchange part of the
interaction. In an optical lattice Mott phases occur,
featuring an interesting behaviour in the magnetic
sector \cite{demler,rizzi}. Experimentally, the ground state of such
gases, its magnetic properties and low temperature dynamical
properties have been studied for $^{23}$Na, which has a spin-singlet ground state
\cite{Ketterle}, and for $^{87}$Rb where the ground state is
ferromagnetic \cite{Sengstock,Chapman,Stamper-Kurn}. The creation of a
one-dimensional (1D) regime for quantum gases by tightly confining the
motion of particles in two directions offered new possibilities for
studying macroscopic quantum effects, and a number of advanced
experiments have been done for spinless 1D bosons
\cite{Pitaevskii,Gora2004}. 

In this paper we consider a 1D liquid of spin-1 bosons with a
short-range interaction containing a repulsive density-density term
and a magnetic exchange interaction. We demonstrate that for the case
of antiferromagnetic exchange the low energy effective action decouples
into two parts: the standard U(1) Tomonaga-Luttinger action
describing density fluctuations, and the action of the O(3) nonlinear
sigma model describing the spin sector. The latter model has a singlet
ground state separated from the first excited triplet by a
gap. Therefore the system of  spin-1 bosons interacting via
antiferromagnetic exchange constitutes a perfect spin liquid where
spin-spin correlations decay exponentially at large distances.  We
also identify the ``order parameter field'', which has correlations
exhibiting a power law decay at T=0. This field consists of 
pairs of bosons
\be 
\Delta = \Psi_1\Psi_{-1} - 1/2\Psi_0^2.
\ee
Hence the fluctuation superfluidity of spin-1 bosons is really a
superfluidity of pairs. 

We derive our results by two means. For weak interactions we
perform a semiclassical analysis of the bosonic path integral. For a
special point where the coupling constants for the density-density and 
exchange interaction are equal to each other we study the Bethe ansatz
solution. The latter result is of a certain interest since it turns
out that the integrable  model of spin-1 bosons may serve as an
integrable regulator for the O(3) nonlinear sigma model.

\section{Low-energy effective action. Semiclassical derivation. }
Our starting point is a three-component Bose gas with Hamiltonian density
\bea
{\cal H}=\frac{1}{2m}\partial_x{\bm\Psi}^\dagger
\partial_x{\bm\Psi}-\mu{\bm\Psi}^\dagger{\bm\Psi}
+\frac{g_0}{2}\left[{\bm\Psi}^\dagger{\bm\Psi}\right]^2
+\frac{g_1}{2}\left[{\bm\Psi}^\dagger{\bm S}^\alpha{\bm\Psi}\right]
\left[{\bm\Psi}^\dagger{\bm S}^\alpha{\bm\Psi}\right]\label{H}.
\eea
Here ${\bm\Psi}(x)=(\Psi_1(x),\Psi_0(x),\Psi_{-1}(x))$ is a
three-component Bose field and ${\bm S}^\alpha$ are spin-1
matrices. We now introduce a number-phase representation 
\bea
\Psi_\sigma(x)=\sqrt{\rho(x)}n_\sigma(x)e^{i\phi_\sigma(x)},
\eea
with a real unit vector field $n_\sigma(x)$
\be
\sum_\sigma n_\sigma^2=1.
\label{constraint}
\ee
In the number-phase representation the Hamiltonian density takes the
form ${\cal H}={\cal H}_0+V$, where
\bea
{\cal  H}_0&=&\frac{1}{8m\rho}(\partial_x\rho)^2
+\frac{\rho}{2m}\sum_\sigma
n_\sigma^2(\partial_x\phi_\sigma)^2+(\partial_x
n_\sigma)^2,\label{kin}\\
V&=&-\mu\rho+\frac{g_0}{2}\rho^2+\frac{g_1}{2}\rho^2\Bigl[(n_1^2-n_{-1}^2)^2
+2n_0^2(n_1^2+n_{-1}^2)\nn
&&\qquad\qquad\qquad+4n_0^2n_1n_{-1}\cos(2\phi_0-\phi_1-\phi_{-1})
\Bigr].
\eea
In order to minimize the potential we eliminate $n_0$ via the
constraint (\ref{constraint}) and define
\be
\phi_\pm=\frac{\phi_1\pm\phi_{-1}}{2}\ ,\qquad
\tilde{\phi}=2\phi_0-\phi_1-\phi_{-1},\qquad
n_\pm=n_1\pm n_{-1}.
\ee
The new phase fields are taken to have ranges
\be
0\leq\phi_+<\pi\ ,\quad 0\leq\phi_-<2\pi\ .
\label{ranges}
\ee
In the new variables the potential takes the form
\bea
V&=&\frac{g_1}{2}\rho^2\left[n_+^2n_-^2
+(2-n_+^2-n_{-}^2)(n_+^2-\frac{n_+^2-n_-^2}{2}[1-\cos\tilde{\phi}])
\right]\nn
&&-\mu\rho+\frac{g_0}{2}\rho^2
 \label{V}.
\eea
Minimizing the potential (\ref{V}) gives the solutions (for $g_1>0$)
\be
A:\quad\rho=\frac{\mu}{g_0}\equiv \rho_0\ ,\quad
\tilde{\phi}=0\ ,\quad n_+=0.
\ee
\be
B:\quad\rho=\frac{\mu}{g_0}\equiv \rho_0\ ,\quad
\tilde{\phi}=\pi\ ,\quad n_-=0.
\ee

From now on we concentrate on solution A. Expanding the potential
around the minimum gives
\be
V=\frac{g_0}{2}(\rho-\rho_0)^2+\frac{g_1\rho_0^2}{2}
\left[2n_+^2+\frac{1}{2}n_-^2(1-\frac{n_-^2}{2})(\tilde\phi)^2\right]+\ldots
\ee
where the dots stand for terms of higher order in powers of $n_{\pm},
\tilde\phi$. Carrying out the analogous expansion for the kinetic
energy part (\ref{kin}) of the Hamiltonian density gives
\bea
{\cal H}_0&=&\frac{\rho_0}{2m}\left[
(\partial_x\phi_+)^2+\frac{n_-^2}{2}(\partial_x\phi_-)^2
+(\partial_xn_-)^2
\frac{1}{2-(n_-)^2}\right]\nn
&&+\frac{1}{8m\rho_0}(\partial_x\rho)^2+\ldots\ .
\eea
Putting everything together we arrive at the following expression for
the action
\bea
S&=&\int dtdx \left[i{\bm \Psi}^\dagger\partial_t{\bm\Psi}-{\cal
    H}\right]
\nonumber\\
&=&\int dtdx\left[-\rho\partial_t\phi_+-\rho_0n_+n_-\partial_t\phi_-
-\frac{\rho_0}{2}\tilde{\phi}n_-\partial_tn_-+\ldots-{\cal H}\right]
\eea
We observe that $\rho$, $n_+$ and $\tilde{\phi}$ correspond to massive
degrees of freedom. In the next step we integrate these out, which in
our quadratic approximation can be done by
``completing the squares''. At energies well below the gaps for
excitations in the $\rho$ $n_+$ and $\tilde{\phi}$ sectors, retardation
effects can be neglected and the low-energy effective action remains
local in time. Integrating over $(\rho-\rho_0)$ generates a contribution 
\be
\frac{1}{2g_0}(\partial_t\phi_+)^2,
\ee
to the action, while integrating out $n_+$ and $\tilde\phi$ generate
respectively 
\be
\frac{1}{4g_1}n_-^2(\partial_t\phi_-)^2
\ee
and
\be
\frac{1}{4g_1}(\partial_tn_-)^2\frac{1}{1-\frac{n_-^2}{2}}.
\ee
Finally we parametrize
\be
n_-=\sqrt{2}\sin\theta ,
\label{theta}
\ee
and arrive at the following form for the low-energy effective action  
$S=S_0+S_{nl\sigma m}$:
\be
S_0=\int dt
dx\left[\frac{1}{2g_0}(\partial_t\phi_+)^2
-\frac{\rho_0}{2m}(\partial_x\phi_+)^2 \right].
\ee
\bea
S_{nl\sigma m}&=&\int dt dx\Bigl\{
\frac{1}{2g_1}\left((\partial_t\theta)^2+\sin^2\theta
(\partial_t\phi_-)^2\right)\nn
&&\qquad\qquad-
\frac{\rho_0}{2m}\left[(\partial_x\theta)^2+\sin^2\theta
(\partial_x\phi_-)^2\right]\Bigr\}.\label{nln}
\eea

The action $S_0$ describes a free massless boson whereas $S_{nl\sigma m}$ is a
parametrization for the O(3) nonlinear sigma model. Indeed, 
defining a three-dimensional unit vector field ${\bm m}^2=1$ by
\be
{\bm m}=
\pmatrix{
\sin\theta\cos\phi_-\cr \sin\theta\sin\phi_-\cr\cos\theta
},
\ee
we obtain
\be
S_{nl\sigma m}=\frac{1}{2g_1}\int dt dx
\left[(\partial_t {\bm m})^2-v_s^2(\partial_x {\bm m})^2\right],
\label{sigmamodel}
\ee
where the spin velocity is given by
\be
v_s=\sqrt{\frac{\rho_0g_1}{m}}.
\ee
The action in the $\phi_+$ sector can be brought to a more standard
Luttinger liquid form
\be
S_0=\frac{K_c}{2\pi}\int dt
dx\left[\frac{1}{v_c}(\partial_t\phi_+)^2
-v_c(\partial_x\phi_+)^2 \right],
\label{boson}
\ee
where the charge velocity $v_c$ and Luttinger parameter $K_c$ are
given by
\be
v_c=\sqrt{\frac{\rho_0g_0}{m}}\ ,\quad
K_c=\pi\sqrt{\frac{\rho_0}{mg_0}}.
\label{Kc}
\ee
We note that for equal coupling constants $g_0=g_1$ the spin and
charge velocities coincide, i.e. $v_s=v_c$.
\subsection{Integration Measure}
In the above derivation we have disregarded the integration measure in the path
integral of our bosonic theory. We still have to verify that it
correctly produces the integration measure for the nonlinear sigma
model (\ref{nln}). We start by considering the Jacobian of the
number-phase parametrization
\be
{\rm det}\frac{\partial(\Psi_\sigma^\dagger,\Psi_\sigma)}
{\partial(\rho,n_1,n_{-1},\phi_\sigma)}=
-4in_1n_{-1}\rho^2.
\ee
Changing variables to $n_\pm$ gives a factor
\be
\frac{1}{2}(n_-^2-n_+^2)dn_+dn_-
\ee
in the integration measure. Expanding $n_+$ around the minimum of the
potential $V$ leaves
\be
n_-^2dn_-=2^{3/2}\sin^2\theta \cos\theta d\theta
\ee
The integration measure then looks:
\be
\sqrt{2}\ d\rho\ d\phi_+\ d\phi_-\ dn_+\ d\theta\ d\tilde{\phi}\ \sin^2\theta
\cos\theta . 
\ee
When integrating out the $\tilde{\phi}$ field we wish to change
variables to
\be
\tilde{\phi}\ n_-\sqrt{1-\frac{n_-^2}{2}}=\hat\phi.
\ee
In terms of the $\theta$-parametrization (\ref{theta}) this gives
\be
d\tilde{\phi}=\frac{d{\hat\phi}}{\sqrt{2}\sin\theta\cos\theta},
\ee
which turns the measure into
\be
d\rho\ dn_+\ d\hat{\phi}\ d\phi_+\left[\sin\theta d\theta\ d\phi_-\right].
\ee
The $\theta,\phi_-$ piece is indeed the correct integration measure
for the nonlinear sigma model.

We conclude that at low energies the spinor-Bose Hamiltonian (\ref{H})
is described by a spin-charge separated theory of a free boson
(\ref{boson}) describing collective superfluid pairing fluctuations
and the O(3) nonlinear sigma model (\ref{sigmamodel}) describing spin
excitations. 

\section{Bethe ansatz solution}
We now turn to an entirely different analysis of the model (\ref{H}),
which will lead to the same conclusion. The first quantized form of the
Hamiltonian (\ref{H}) is
\bea
H = -\frac{1}{2m}\sum_{j=1}^N\frac{\p^2}{\partial x_j^2} +
\frac{1}{2}\sum_{i\neq j}\delta(x_i-x_j)[g_0 + g_1{\bf S}_i{\bf
    S}_j] ,
\label{model} 
\eea
where ${\bf S}$ are spin $S=1$ operators. It was shown in
Ref. \cite{cao} that the model (\ref{model}) is
integrable along the line $g_1 = g_0 $. A simple understanding of
certain features of the exact solution can be gained by considering
the case of two bosons in the strong coupling limit (in which we
neglect the kinetic energy). The potential energy of the pair of
bosons at the same position is obtained by adding two spins $S=1$ to
form total angular momentum $J$: 
\be
E_J = g_0 + g_1[\frac{1}{2}J(J+1) - S(S+1)],\quad S=1,\ J =0,1,2.
\ee
The singlet $J=0$ is always lower in energy than the triplet and
quintet states and at $g_1 > g_0/2$ its energy becomes negative. 
In this regime the formation of $J=0$ bound states becomes
energetically favourable compared to having scattering states of
two bosons. As a result the ground state is made of paired bosons. The
integrable case $g_0=g_1$ lies well within the parameter region where
this occurs.

Let us now return to the Bethe ansatz analysis. According to
Ref.\cite{cao} the Bethe ansatz equations are given by
\bea
\re^{\ri k_i L} &=& -\prod_{j=1}^N e_4(k_i -k_j)\prod_{a=1}^M
e_{-2}(k_i - \lambda_a)\ ,\nonumber\\ 
-1&=&\prod_{j=1}^Ne_{-2}(\lambda_a - k_j) \prod_{b=1}^Me_2(\lambda_a -
\lambda_b)\ ,\nonumber\\ 
E &=& \frac{1}{2m}\sum_{i=1}^Nk_i^2\ ,\quad
P=\sum_{i=1}^Nk_i\ ,
\label{bethe}
\eea
where $N$ is the total number of particles, $M = (N - S^z)$, $E$ and
$P$ are respectively the energy and momentum and
\be
e_n(x) = \frac{x + \ri cn/2}{x - \ri cn/2}.
\ee
The parameter $c$ is related to the couplings $g_{0,1}$ by
\be
c=mg_0=mg_1.
\ee
In order to proceed, we now assume that the system is at a finite
temperature $T\ll\mu$, where the chemical potential $\mu$ is the
largest energy scale in the problem. As was shown in
Ref.~\cite{cao}, in this case \cite{TsWi83} only three
types of (string) solutions of the Bethe ansatz equations
(\ref{bethe}) contribute to the thermodynamics of the model
\begin{itemize}
\item
Bound states of $k$'s and $\lambda$'s ($k$-$\Lambda$ strings
\cite{Takahashi72,Hubbardreview,book}): 
\bea
k^{\pm}_i &=& \Lambda_i \pm \ri c/2 + \gamma^{\pm}_i\ , \nonumber\\
\lambda^{\pm}_i &=& \Lambda_i \pm \ri c/2 + \delta^{\pm}_i\ ,
\label{bound}
\eea
where the string centres $\Lambda_i$ are real, $\gamma, \delta
\propto\exp(-{\rm const}\ L)$. The zero temperature ground state is
formed by a partially filled Fermi sea of these bound 
states. 
\item
Real $k$'s.
\item
$\lambda$-strings\cite{Takahashi71a,Takahashi72,Hubbardreview,book}: 
\bea
\lambda_a^{(n,j)} = \lambda^{(n)}_a + \ri c[(n+1)/2 -j] + O(\exp(-L
       {\cal N})), ~~ j =1,...n \ ,
\eea
where the string centres $\lambda^{(n)}_a\in\mathbb{R}$ and 
${\cal N}$ is related to the number of excitations in this sector.
\end{itemize}

Substituting (\ref{bound}) into (\ref{bethe}) and dropping the
exponentially small corrections $\gamma, \delta$ we obtain the
following set of equations
\bea
\re^{2\ri L\Lambda_i} &=& \prod_{j=1}^{N_b}{\cal E}(\Lambda_i -
\Lambda_j)\prod_{p=1}^{N_r}{\cal P}(\Lambda_i - k_p)\ ,\label{b2}\\ 
\re^{\ri Lk_i} &=& \prod_{j=1}^{N_r}e_4(k_i -
k_j)\prod_{p=1}^{N_p}{\cal P}(k_i -
\Lambda_p)\prod_{a=1}^{M_r}e_{-2}(k_i - \lambda_a) \ ,
\label{b3}\\
-1&=&\prod_{j=1}^{N_r}e_{-2}(\lambda_a - k_j)
\prod_{b=1}^{M_r}e_2(\lambda_a - \lambda_b)\ ,\label{b4}\\ 
E &=& \frac{1}{2m}\left[\sum_{i=1}^{N_b}(2\Lambda_i^2 - c^2/2) +
  \sum_{i=1}^{N_r}   k_i^2\right]\ ,\quad
P=\sum_{i=1}^{N_b}2\Lambda_i+\sum_{p=1}^{N_r}k_p.
\label{bethe2} 
\eea
Here $N = N_r + 2N_b$, $S^z = N_r - M_r$ and 
\[
{\cal E}(x) = e_6(x)e_4(x)e_{-2}(x), ~~ {\cal P}(x) = e_5(x)e_{-1}(x).
\]
We now take the logarithm of the Bethe
ansatz equations for the $k$-$\Lambda$ strings forming the ground
state and then express them in terms of a counting function
\cite{yaya69,book}
\bea
y(\Lambda)&=&2\Lambda+\frac{1}{L}\sum_{j=1}^{N_b}
\theta\Bigl(\frac{\Lambda-\Lambda_j}{3c}\Bigr)+
\theta\Bigl(\frac{\Lambda-\Lambda_j}{2c}\Bigr)-
\theta\Bigl(\frac{\Lambda-\Lambda_j}{c}\Bigr)\nn
&&+\frac{1}{L}\sum_{p=1}^{N_r}
\theta\Bigl(2\frac{\Lambda-k_p}{5c}\Bigr)-
\theta\Bigl(2\frac{\Lambda-k_p}{c}\Bigr).
\label{countingfn}
\eea
Here $\theta(x)=2\arctan(x)$.
In terms of the counting function the Bethe ansatz equations
(\ref{b2}) read
\be
y(\Lambda_j)=\frac{2\pi I_j}{L}\ ,
\label{yIj}
\ee
where $I_j$ are integer (for $N_b$ odd) of half-odd integer
(for $N_b$ even) numbers. We note that the momentum is expressed in
terms of the $I_j$ as
\be
P=\frac{2\pi}{L}\sum_{i=1}^{N_b}I_j+\sum_{p=1}^{N_r}k_p.
\label{momentum}
\ee
As shown in \cite{cao} the ground state of the system corresponds to
choosing $N_r=M_r=0$, $2N_b=N$ and
\be
I_j=-\frac{N_b+1}{2}+j\ ,\quad j=1,\ldots N_b.
\ee
This immediately tells us that the ``Fermi momentum'' is
\be
k_F=\frac{2\pi}{L}(I_{N_b}+\frac{1}{2})=\frac{\pi N}{2L}=\frac{\pi}{2}\rho_0.
\label{kF}
\ee
We are interested in the limit
\be
L\to\infty\ ,\quad N_b\to\infty\ ,\quad \frac{2N_b}{L}=\rho_0=\ {\rm const}.
\ee
In this limit we can turn the sum over $j$ in (\ref{countingfn}) into
an integral by means of the Euler-Maclaurin sum formula, resulting in
\bea
y(\Lambda)&=&2\Lambda+\int_{-\infty}^\infty d\Lambda'\left[
\theta\Bigl(\frac{\Lambda-\Lambda'}{3c}\Bigr)+
\theta\Bigl(\frac{\Lambda-\Lambda'}{2c}\Bigr)-
\theta\Bigl(\frac{\Lambda-\Lambda'}{c}\Bigr)\right]\sigma'_p(\Lambda')\nn
&&\qquad +\frac{1}{L}\sum_{p=1}^{N_r}
\theta\Bigl(2\frac{\Lambda-k_p}{5c}\Bigr)-
\theta\Bigl(2\frac{\Lambda-k_p}{c}\Bigr).
\label{countingfn2}
\eea
Here $\sigma'_{p,h}(\Lambda)$ is are the root densities of ``particles''
and ``holes'' for $k$-$\Lambda$ strings. They are related to
the counting function by
\be
\frac{dy(\Lambda)}{d\Lambda}=2\pi
\left[\sigma'_p(\Lambda)+\sigma'_h(\Lambda)\right].
\ee
Taking the derivative of (\ref{countingfn2}) we obtain an integral
equation 
\be
\sigma'_p(\Lambda)+\sigma'_h(\Lambda)=\frac{1}{\pi}
+(a_6+a_4-a_2)*\sigma_p'\Bigl|_\Lambda
+\frac{1}{L}\sum_{p=1}^{N_r}a_5(\Lambda-k_p)-a_1(\Lambda-k_p).
\label{gseq}
\ee
Here $*$ denotes a convolution
\be
f*g\Bigl|_x = \int_{-\infty}^{\infty}dy \ f(x-y)g(y).
\ee
and the integral kernels are given by 
\be
a_n(x) = \frac{1}{2\pi}\frac{nc}{x^2 + (nc/2)^2}.
\ee
Equation (\ref{gseq}) can be recast in the form
\be
(I+a_2)*(I-a_4)*\sigma'_p\Bigl|_\Lambda=\frac{1}{\pi}-\sigma'_h(\Lambda)
+\frac{1}{L}\sum_{p=1}^{N_r}a_5(\Lambda-k_p)-a_1(\Lambda-k_p).
\label{inteqsigmap}
\ee
In order to derive the Bethe ansatz equations (\ref{b3}) over the
ground state of the system we need to express 
$\prod_{p=1}^{N_b}{\cal P}(k_i-\Lambda_p)$ in terms of the root
densities for $k$-$\Lambda$ strings. We have
\bea
\prod_{p=1}^{N_b}{\cal P}(k-\Lambda_p)&=&
\exp\left(-i\sum_{p=1}^{N_b}\theta\Bigl(2\frac{k-\Lambda_p}{5c}\Bigr)
-\theta\Bigl(2\frac{k-\Lambda_p}{c}\Bigr)\right)\nn
&=&\exp\left(-iL\int_{-\infty}^\infty d\Lambda\sigma'_p(\Lambda)\left[
\theta\Bigl(2\frac{k-\Lambda}{5c}\Bigr)
-\theta\Bigl(2\frac{k-\Lambda}{c}\Bigr)\right]\right)\nn
&\equiv&\exp\left(2\pi iLf(k)\right).
\label{fofk0}
\eea
Taking derivatives we have
\be
f'(k)=(a_1-a_5)*\sigma_p\Bigl|_k=a_1*(I-a_4)*\sigma_p\Bigl|_k.
\ee
On the other hand Eq.~(\ref{inteqsigmap}) implies that
\be
a_1*(I-a_4)*\sigma_p\Bigl|_k=\frac{1}{2\pi}-G*\sigma'_h\Bigl|_k
-\frac{1}{L}\sum_{p=1}^{N_r}(a_2-a_4)\Bigl|_{\Lambda-k_p},
\ee
where $G=a_1*(I+a_2)^{-1}$. The kernel of $G$ is given by
\be
G(\Lambda)=\frac{1}{2c\cosh\Bigl(\frac{\pi\Lambda}{c}\Bigr)}.
\label{G}
\ee
As a result we have
\bea
f(k)&=&\frac{k}{2\pi}-\frac{1}{2\pi L}
\sum_p\theta\Bigl(\frac{k-k_p}{c}\Bigr)
-\theta\Bigl(\frac{k-k_p}{2c}\Bigr)
-\Gamma*\sigma'_h\Bigl|_k+\chi,
\label{fofk}
\eea
where the $k$-independent piece $\chi$ is fixed by the requirement that
$f(\infty)=0$, and $\Gamma$ is an integral operator with the kernel
\be
\Gamma(x)=\frac{1}{\pi}\arctan\Bigl(\tanh\Bigl(\frac{\pi x}{2c}\Bigr)\Bigr).
\ee
In order to proceed we need to evaluate $\Gamma*\sigma'_h$. For the ground
state this is simply given by
\be
\Gamma*\sigma'_h\Bigl|_k=\int_{{Q_0}}^\infty d\Lambda
\left[\Gamma(\Lambda+k)-\Gamma(\Lambda-k)\right] \sigma'_0(\Lambda)\ ,
\label{sigma0}
\ee
where
\be
\sigma'_0(\Lambda)=\frac{1}{\pi}+\int_{-{Q_0}}^{Q_0}
d\Lambda'\ K(\Lambda-\Lambda')\
\sigma_0'(\Lambda').
\label{rootdensGS}
\ee
Here we have defined a kernel
\be
K=a_6+a_4-a_2.
\ee
Using the relations $\sigma_0(\Lambda)=\sigma_0(-\Lambda)$,
$\lim_{\Lambda\to\infty}\sigma_0(\Lambda)=\frac{1}{\pi}$ and 
taking into account that by
virtue of Eq.~(\ref{sigma0}) the quantity $\Gamma*\sigma'_h\Bigl|_k$ is an odd function
of $k$ we conclude that for the ground state configuration we have $\chi=0$.
Let us now consider the limit where
\be
c,|k|\ll {Q_0}.
\ee
For large $x$ we have
\be
\Gamma(x)\longrightarrow\frac{1}{4}-\frac{1}{\pi}e^{-\pi x/c}\ ,\quad
x\gg c,
\ee
which yields
\be
\Gamma*\sigma'_h\Bigl|_k\approx\frac{2}{\pi}\sinh\Bigl(\frac{\pi k}{c}\Bigr)
\int_{{Q_0}}^\infty d\Lambda
e^{-\pi\Lambda/c}\ \sigma'_0(\Lambda)\equiv \frac{\Delta}{2\pi}
 \sinh\Bigl(\frac{\pi k}{c}\Bigr).
\label{gammasigma}
\ee
For an excited state with $n_p$ particles and $n_h$ holes with
corresponding spectral parameters $\Lambda^{p,h}_j$ added to the
ground state distribution of $k$-$\Lambda$ strings as well as $N_r$
additional real $k$'s, the integral equation for the total root density
$\sigma'=\sigma'_p+\sigma'_h$ reads
\bea
&&\sigma'(\Lambda)=\frac{1}{\pi}+\int_{-Q}^Qd\Lambda'\
K(\Lambda-\Lambda')\ \sigma'(\Lambda')\nn
&&+\frac{1}{L}\left[\sum_{l=1}^{n_p}K(\Lambda-\Lambda^p_l)
-\sum_{r=1}^{n_h}K(\Lambda-\Lambda^h_r)
+\sum_{j=1}^{N_r}(a_5-a_1)(\Lambda-k_j)\right].
\label{sigmaex}
\eea
Here the integration boundary $Q$ may differ from the one for the
ground state ($Q_0$) by a contribution of at most order ${\cal O}
(L^{-1})$. The integral equation (\ref{sigmaex}) can be solved by an
expansion in inverse powers of $L$
\be
\sigma'(\Lambda)=\sigma'_0(\Lambda)+\frac{1}{L}\sigma'_1(\Lambda)+{\cal O}(L^{-2}),
\ee
where $\sigma_0$ is the root density or the ground state
(\ref{sigma0}). Hence the leading contribution (in $1/L$) to
$\Gamma*\sigma'_h\Bigl|_k$  is given by (\ref{gammasigma}). The
contribution due to $\sigma'_1$ can be seen to be of the form
\be
\frac{1}{L}\left[{\rm const}+{\cal O}\Bigl(\exp\left[-\frac{\pi
      Q}{c}\right]\Bigr)\right],
\quad c,|k|\ll Q.
\ee
The constant (i.e. independent of $k$) contribution is by construction
precisely cancelled by $\chi$ in (\ref{fofk}), so that

\bea
f(k)&=&\frac{k}{2\pi}
-\frac{\Delta}{2\pi}
 \sinh\Bigl(\frac{\pi k}{c}\Bigr)
-\frac{1}{2\pi L}
\sum_{p=1}^{N_r}\theta\Bigl(\frac{k-k_p}{c}\Bigr)
-\theta\Bigl(\frac{k-k_p}{2c}\Bigr)\nn
&&
+\frac{1}{L}{\cal O}\Bigl(\exp\left[-\frac{\pi
      Q}{c}\right]\Bigr).
\label{fofk2}
\eea
Substituting this into (\ref{fofk}), (\ref{fofk0}) and dropping the
exponentially small (in $Q/c$) contribution leads to
\bea
\prod_{p=1}^{N_b}{\cal P}(k-\Lambda_p)&=&
e^{iLk}\ e^{-iL\Delta \sinh(\pi k/c)}
\prod_{p=1}^{N_r}e_{-4}(k-k_p)e_2(k-k_p).
\eea
Using this in the Bethe ansatz equations (\ref{b3}) and then
rescaling variables by
\be
k = \frac{c\theta}{\pi}\ ,\quad \lambda = \frac{c u}{\pi},
\ee
we arrive at the Bethe ansatz equations for the O(3)
nonlinear sigma model \cite{Wieg, AMTsigma}: 
\bea
&& \re^{\ri \Delta L\sinh\theta_l} = \prod_{j=1}^{N_r}\frac{\theta_l -
  \theta_j + \ri\pi}{\theta_l - \theta_j -
  \ri\pi}\prod_{a=1}^{M_r}\frac{\theta_l - u_a - \ri\pi}{\theta_l -
  u_a + \ri\pi}\ ,\quad l=1,\ldots N_r,\\ 
&& \prod_{j=1}^{N_r}\frac{u_a - \theta_j + \ri\pi}{u_a - \theta_j -
  \ri\pi}=- \prod_{b=1}^{M_r}\frac{u_a - u_b + \ri\pi}{ u_a - u_b -
  \ri\pi} \ ,\quad a=1,\ldots, M_r.
\label{baesigma}
\eea
The contribution of spin excitations to the total momentum is given by
\be
P_{\rm spin}=\Delta\sum_{l=1}^{N_r}\sinh\theta_l.
\ee
We note that in addition to (\ref{baesigma}) we still have
``physical'' Bethe ansatz equations that determine the rapidities of
the particle and hole excitations over the sea of $k$-$\Lambda$
strings forming the ground state. It follows from the finite-volume
quantization conditions (\ref{baesigma}) that the dressed phase-shifts
for scattering of such particles and holes with excitations in the
spin sector vanish (up to exponentially small corrections in $Q/c$).
Hence the quantization conditions in the gapless sector must be of the
form
\bea
e^{iLP(\Lambda^p_j)}&=&\prod_{k\neq
  j}^{n_p}S_{pp}(\Lambda^p_j-\Lambda^p_k)
\prod_{m=1}^{n_h}S_{ph}(\Lambda^p_j-\Lambda^h_m)\ ,\ j=1,\ldots,n_p\ ,\nn
e^{iLP(\Lambda^h_l)}&=&\prod_{k=1}^{n_p}
S_{hp}(\Lambda^h_l-\Lambda^p_k)
\prod_{m\neq l}^{n_h}S_{hh}(\Lambda^h_l-\Lambda^h_m)\ ,\ l=1,\ldots,
n_h.
\eea
Here $P(\Lambda)$ is the dressed momentum for $k$-$\Lambda$ strings
and $S_{ab}(\Lambda)$ the dressed scattering phases for particles and
holes. They can be expressed in terms of the solution to the integral
equation (\ref{sigmaex}) following standard methods \cite{smat,book}.
\subsection{Dressed Energies}
We will now show that the energy in the spin sector is given by
\bea
E_{\rm spin} = \Delta v_c\sum_{l=1}^{N_r} \cosh\theta_l\ .
\eea
Our starting point are the equations for the dressed energies for
$k$-$\Lambda$ strings $\eps'(\Lambda)$ and real $k$'s
$\kappa(k)$\cite{cao} 
\bea
\eps'(\Lambda)&=&\frac{\Lambda^2}{m}-\frac{c^2}{4m}-2\mu+\int_{-Q_0}^{Q_0} d\Lambda'\ 
K(\Lambda -\Lambda')\ \eps'(\Lambda')\ ,\nn
\kappa(k)&=&\frac{k^2}{2m}-\mu+\int_{-Q_0}^{Q_0}d\Lambda\ 
(a_5-a_1)(k-\Lambda)\ \eps'(\Lambda).
\eea
Here $\mu$ is the chemical potential related to $Q_0$ by the
requirement
\be
\eps'(Q_0)=0.
\ee 
Using the integral equation for $\eps'$ in the equation for $\kappa$
we find that 
\be
\kappa(k)=\int_{Q_0}^{\infty}d\Lambda\left[G(k-\Lambda)+G(k+\Lambda)\right]
\eps'(\Lambda),
\ee
where $G$ is given in Eq.~(\ref{G}). In the limit $c,|k|\ll Q_0$ this simplifies to
\be
\kappa(k)\approx\cosh\Bigl(\frac{\pi k}{c}\Bigr)\ \frac{2}{c}
\int_{Q_0}^{\infty}d\Lambda\ e^{-\pi\Lambda/c}\ \eps'(\Lambda).
\ee
Expanding $\eps'$ around $Q_0$ we have
\be
\eps'(\Lambda)=\frac{d\eps'}{d\Lambda}\Bigl|_{Q_0}(\Lambda-Q_0)+\ldots
\ee
Noting that the Fermi velocity in the charge sector is defined as
\be
v_c=\frac{d\eps'}{dp}\Bigl|_{Q_0}=
\frac{\frac{d\eps'}{d\Lambda}}{\frac{dp}{d\Lambda}}\Bigl|_{Q_0}\ ,
\ee
where $p(\Lambda)$ is the dressed momentum for $k$-$\Lambda$ strings
and using that this may be expressed as $p(\Lambda)=y(\Lambda)$ by
virtue of Eqs.~(\ref{momentum}) and (\ref{yIj}), we conclude that
\be
v_c= \frac{\frac{d\eps'}{d\Lambda}}{2\pi\sigma'(\Lambda)}\Bigl|_{Q_0},
\ee
and thus
\be
\kappa(k)\approx v_c\cosh\Bigl(\frac{\pi k}{c}\Bigr)\ \frac{4c}{\pi}
\sigma'(Q_0)\ e^{-\pi Q_0/c}.
\ee
Evaluating the right-hand side of Eq.~(\ref{gammasigma}) by using an analogous approximation and rescaling $k$
as before we conclude that indeed 
\be
\kappa(\theta)=\Delta v_c\cosh\theta.
\ee
\section{Scaling Dimensions.}
The analysis of the finite-size spectrum is complicated by the fact
that the ground state is made of strings. A priori this could lead to
complications related to deviations of the strings from their ideal
forms (see, e.g. \cite{holger}). However, as the critical sector of our
theory is a simple compactified boson we do not expect such
complications to play a role.
The dressed charge $Z=Z(Q_0)$ is defined in terms of the solution of
the integral equation \cite{vladb}
\bea
Z(\Lambda)&=&1+\int_{-Q_0}^{Q_0}d\Lambda'\ K(\Lambda-\Lambda')\ Z(\Lambda').
\eea
We see that we have
\be
Z=\pi\sigma'(Q_0).
\ee
It is easy to see that the dressed charge only depends on the
dimensionless ratio
\be
\gamma=\frac{c}{\rho_0}.
\ee
At small $\gamma$ we find that
\be
Z\to \frac{\sqrt{\pi}}{2}\gamma^{-1/4}\gg 1.
\ee
On the other hand, for $\gamma\to\infty$ the dressed charge approaches
$1$ from below. This implies that $Z$ is not a monotonic function of
$\gamma$, but instead has a local minimum.
The finite size spectra of energy and momentum are expressed in terms
of the dressed charge as \cite{dimensions}
\bea
E(\Delta N_b,D,N_+,N_-)&=&\frac{2\pi v_c}{L}\left[\frac{(\Delta
    N_b)^2}{4Z^2}+Z^2D^2+N_++N_-\right]\ ,\\ 
P(\Delta N_b,D,N_+,N_-)&=&\frac{2\pi}{L}\left[N_+-N_-+D(\Delta N_b)\right]+2k_FD.
\eea
Here $k_F$ is given by Eq.~(\ref{kF}), $\Delta N_b$ is the change in bound
state number compared to the ground state and $D$ is an integer. The
corresponding spectrum of scaling dimensions is given by
\bea
h(\Delta N_b,D,N_+,N_-)&=&\frac{1}{2}\left[DZ+\frac{(\Delta
    N_b)}{2Z}\right]^2+N_+\ ,\nn
\bar{h}(\Delta N_b,D,N_+,N_-)&=&\frac{1}{2}\left[DZ-\frac{(\Delta
    N_b)}{2Z}\right]^2+N_-.
\label{dimsBA}
\eea
Let us look at some specific examples. The lowest charge-neutral
excitations with momenta $2k_F$ and $0$ respectively correspond to
$\Delta N_b=N_\pm=0$, $D=1$ and  $D=N_b=N_-=0$, $N_+=1$ 
\bea
h(0,1,0,0)&=&\frac{Z^2}{2}\ ,\nn
h(0,0,1,0)&=&1.
\eea
The lowest excited state with charge $2$ corresponds to $\Delta N_b=1$,
$N_\pm=D=0$ 
\be
h(1,0,0,0)=\frac{1}{8Z^2}.
\ee

\section{Correlation functions.}
Let us now turn to the large-distance asymptotics of correlation
functions. We start by considering the gapless sector of the
low-energy effective action, which is given by
\be
S_0=\frac{K_c}{2\pi}\int dt
dx\left[\frac{1}{v_c}(\partial_t\phi_+)^2
-v_c(\partial_x\phi_+)^2 \right].
\label{boson2}
\ee
Given that the original phase fields $\phi_{1,2}$ were
$2\pi$-periodic, the field $\phi_+$ is in fact $\pi$-periodic (see
Eq.~(\ref{ranges})). As a result, the spectrum of scaling dimensions of
primary operators in the Gaussian model (\ref{boson2}) 
is given by 
\be
h(m,n)=\frac{1}{8}\left(\frac{2m}{\sqrt{K}}+n\sqrt{K}\right)^2\ ,\qquad
\bar{h}(m,n)=\frac{1}{8}\left(\frac{2m}{\sqrt{K}}-n\sqrt{K}\right)^2\ ,
\label{dims}
\ee
where $n$ and $m$ are integers. We note that (\ref{dims}) agrees with
(\ref{dimsBA}) in the appropriate limit, where we have $\sqrt{K}=2Z$.
The operators corresponding to these scaling dimensions are
constructed by introducing the field dual to $\phi_+$ by\cite{haldane}
\be
[\theta_+(x),\phi_+(x')]=i\frac{\pi}{2}\sgn(x-x').
\ee
The boson number is given by
\be
N=\frac{1}{\pi}\int dx\ \partial_x\theta_+(x),
\ee
The local operators corresponding to the spectrum of scaling
dimensions (\ref{dims}) are
\be
{\cal O}_{2m,n}=e^{2im\phi_++in\theta_+}\ ;\quad m,n\ {\rm integer}.
\ee
These operators carry charge $2m$, reflecting the fact that our
fundamental objects are pairs of bosons and have two-point functions
of the form\footnote{We note that
if $\phi_+$ were a $2\pi$-periodic field, the allowed scaling
dimensions would be
$
h_{2\pi}(m,n)=\frac{1}{8}\left(\frac{m}{\sqrt{K}}+2n\sqrt{K}\right)^2$,
$\bar{h}_{2\pi}(m,n)=\frac{1}{8}\left(\frac{m}{\sqrt{K}}-2n\sqrt{K}\right)^2$
and correspond to operators ${\cal O}_{m,2n}$. This is the case for
Haldane's ``bosonization of the boson''\cite{haldane}.}
\be
\langle{\cal O}_{2m,n}(x){\cal  O}^\dagger_{2m,n}(0,0)\rangle
=\left[\frac{a_0}{x}\right]^{2h(m,n)+2\bar{h}(m,n)}
e^{in\pi\rho_0x},
\ee
where $a_0$ is a short-distance cutoff
determined by the gapped degrees of freedom we have integrated out in
order to arrive at the action (\ref{sigmamodel}), (\ref{boson}).
We are now in a position to
write down expressions for operators that do not involve the spin
sector. These are the pairing field $\Delta$ and the density, which
are given by 
\bea
\Delta &=& \Psi_1\Psi_{-1} - \Psi_0^2/2 \ ,\nn
:\rho: &=& \Psi^+_{\sigma}\Psi_{\sigma} - \rho_0 \ .
\label{rho1}
\eea
As $\Delta$ carries charge $-2$ we conclude that it takes the form
\be
\Delta(x)=\rho_0\sum_{n=-\infty}^\infty A_n {\cal O}_{2,n}(x)\ ,
\label{delta}
\ee
where $A_n$ are numerical coefficients. We will see below that
$A_{2k+1}=0$. Similarly, as the density is neutral it is given by 
\be
:\rho:=\frac{1}{\pi}\partial_x\theta_++
\sum_{n=-\infty}^\infty B_n {\cal O}_{0,n}(x)\ .
\ee
Like for the pairing field, it turns out that by virtue of the
statistics of the Bose field all odd coefficients vanish,
i.e. $B_{2k+1}=0$.  
Let us now turn to the Bose field itself. At low energies the
semiclassical analysis gives
\be
\Psi_\sigma(x)\sim \sqrt{\rho_0}\re^{i\phi_+}
\pmatrix{
\frac{m_1+im_2}{\sqrt{2}}\cr
m_3\cr
-\frac{m_1-im_2}{\sqrt{2}}\cr}
.\label{Psi}
\ee
Since the spin part of the bosonic operator is proportional to the
slowly varying unit vector field ${\bf m}$, we conclude that the
triplet excitations of the nonlinear sigma model occur at small
momenta $q\approx 0$. To take into account the quantization of total
charge we should complete Eq.~(\ref{Psi}) with a piece containing the field dual
to $\phi_+$, which amounts to the substitution
\bea 
\sqrt{\rho_0}{\re}^{i\phi_+} \longrightarrow  
\sqrt{\rho_0+\frac{1}{\pi}\partial_x\theta_+}
\sum_n C_{2n} e^{2in\theta_++2in\pi\rho_0x}\re^{i\phi_+}\ .
\label{Psi2}
\eea
We note that only even exponentials of the dual field are allowed by
virtue of the bosonic statistics of the fields $\Psi_\sigma$, which
requires the Lorentz spin of the operators appearing in Eq.~(\ref{Psi2})
to be integer. At $T=0$ we then obtain
\bea
\langle T_\tau\Psi_\sigma^\dagger(\tau,x)\Psi_{\sigma'}(0,0)\rangle&=&
\delta_{\sigma,\sigma'}\ \langle T_\tau
m_\sigma(\tau,x)\ m_\sigma(0,0)\rangle f_{\rm charge}(\tau,x)\ ,
\eea
\bea
&&f_{\rm charge}(\tau,x)=
\rho_0\left[\frac{a_0^2}{(v^2\tau^2 + x^2)}\right]^{1/4K_c} \nn
&&+ \rho_0\left[\frac{a_0^2}{(v^2\tau^2 + x^2)}\right]^{K_c+1/4K_c}
\left[C_2^2e^{2\pi i\rho_0 x}\frac{v\tau+ix}{v\tau-ix}+{\rm h.c.}\right]+\ldots .
\eea
The correlation function of the sigma model is well approximated by the
expression \cite{balog} 
\be
\langle T_\tau m_\sigma(\tau,x)\ m_\sigma(0,0)\rangle
\approx ZK_0(\Delta\sqrt{\tau^2+x^2/v^2}),
\ee
where $Z$ is a normalization. We note that finite temperature
correlation functions can be calculated by using the conformal mapping
in the gapless sector and the method of Ref.~\cite{EK08} for the
nonlinear sigma model.

The expression for the Bose-fields feed back to the low-energy
projections of the density and pairing field, which are bilinears in
the Bose fields. Given that only even bosonic exponentials of the dual
field appear in the expression for $\Psi_\sigma$, the same must hold
for $:\rho:$ and $\Delta$. This requirement sets the coefficients of
${\cal O}_{2n,2k+1}$ in the expressions for $:\rho:$ and $\Delta$ to
zero. The two-point functions of both these operators undergo a power law
decay at large distances
\bea
\langle\Delta(x)\Delta^\dagger(0)\rangle&\sim&\rho_0^2|A_0|^2
\left[\frac{a_0^2}{x^2}\right]^{1/K_c} \nn
&&+2|A_2|^2\cos(2\pi\rho_0x)
\left[\frac{a_0^2}{x^2}\right]^{K_c+1/K_c}+\ldots \ ,
\eea
\bea
\langle:\rho(x):\ :\rho(0):\rangle&\sim&
\frac{K_c}{2\pi^2x^2}+2B_2^2\cos(2\pi\rho_0x)
\left[\frac{a_0^2}{x^2}\right]^{K_c}+\ldots .
\eea
We see that the oscillating piece of the two-point function of the
pairing field $\Delta$ always decays more rapidly with distance than
the oscillating piece of the charge density.

\subsection{Optical Lattice}
The model (\ref{H}) arises as the low-density continuum limit of a
three-component Bose-Hubbard system
\be
H_{BH}=-t\sum_{\sigma}a^\dagger_{j,\sigma}a_{j+1,\sigma}+{\rm
  h.c.}+U_0\sum_j n_j^2+U_1\sum_j {\bm S}_j^2-\mu\sum_j n_j\ ,
\label{BH}
\ee
where $n_j=\sum_\sigma a^\dagger_{j,\sigma}a_{j,\sigma}$ and
$S_j^\alpha$ are spin-1 operators on site $j$. An interesting question
is what the phase diagram of the model (\ref{BH}) looks like.
A Density Matrix Renormalization Group analysis
\cite{rizzi} has established the regions in parameter space where Mott
phases occur. 
In order to address this issue by analytical methods one would need to
bosonize the Hamiltonian (\ref{BH}) for strong interactions and large
densities. Assuming that the operator content suggested by Eqs.~(\ref{Psi}) 
and (\ref{Psi2}) remains unchanged, this would occur when the
oscillating part of the Bose field becomes sufficiently relevant so
that the oscillating parts of the operator $n_j^2$ turn into a
relevant perturbation in the entire charge sector. The perturbation is
of the form 
\be
\delta S=\lambda\int dtdx\ {\cal O}_{0,4}(t,x).
\ee
It is relevant in the RG sense if $K_c<\frac{1}{2}$. It would be
interesting to check whether the phase boundaries established in
\cite{rizzi} coincide with the loci in parameter space where $K_c$
becomes $1/2$.

\section{Discussion.}
In this work we have analyzed a continuum model of a spinor Bose gas
in one spatial dimension. Using standard semiclassical methods we
showed that for weak interactions the low energy degrees of freedom
are described by a spin-charge separated theory of a free boson and
the O(3) nonlinear sigma model. We then turned to the integrable line
in the model. From the exact Bethe ansatz solution we obtained an
independent derivation of the low-energy effective theory. An
interesting by-product of our analysis is the demonstration that the
integrable model provides a simple, integrable regularization of the
O(3) nonlinear sigma model. Finally, we determined the long-distance
asymptotics of certain correlation functions (including the
``order-parameter'' correlator) in the framework of the low-energy
description. The dominant fluctuations in the theory are of
superfluid singlet pair type. Our analysis for dynamical correlations
pertains mostly to the weak-coupling regime. The opposite limit of
(infinitely) strong repulsive density-density interaction should
amenable to a treatment along the lines set out in \cite{glaz}.
Other interesting open questions are how the 1D spinor Bose gas
behaves under a quantum quench \cite{austen} and what happens in the
unbalanced case where the densities of the three species of bosons are
fixed at different values. Finally, it would be interesting to
generalize our analysis to spin $S$ bosons \cite{amico}.

\ack
We thank D. Kovrizhin, P. Lecheminant, J.E. Moore and particularly D. Schuricht for
helpful discussions. The work was supported by the EPSRC under grant
EP/D050952/1 (FHLE), the US DOE under contract number DE-AC02-98 CH
10886 (AMT), the IFRAF Institute (AMT), the ANR through grants
05-BLAN-0205 and 06-NANO-014-01 (GS), by the ESF networks INSTANS and
QUDEDIS, and by the Dutch Foundation FOM (GS). FHLE and AMT  thank the
Galileo Galilei Institute for Theoretical Physics for kind
hospitality and INFN for partial support during the completion of this
work.

\end{document}